\documentclass{elsarticle}

\usepackage{graphicx,subfigure,bm,amsmath,amssymb}
\begin{document}

\title{Information Sharing for Strong Neutrals on Social Networks - Exact Solutions for Consensus Times}

\author{Chjan Lim\fnref{lim}}
\author{William Pickering\fnref{pickering}}
\fntext[lim]{limc@rpi.edu}
\fntext[pickering]{pickew@rpi.edu}

\address{NEST, Rensselaer Polytechnic Institute,110 8th Street,
Troy, New York 12180¨C3590, USA} 
\address{Department of Mathematical Sciences Rensselaer Polytechnic Institute, 110 8th
Street, Troy, New York 12180-3590, USA}




\begin{abstract}
To analyze the nuances of the root concept of neutral in social networks,
we focus on several related interpretations and suggest corresponding
mathematical models for each of them from the family of
information-sharing multi-agents network games known as Voter models and the Naming
Games (NG). We solve the case of the strong neutrals known as the middle-roaders for global quantities
such as expected times to consensus and local times. By using generating functions and treating the two extreme and middle opinions in this modification
as a two balls, three urns version of the Voter model, we give closed-form expressions for the eigenvalues and eigenvectors
of its Markov propagator. This modification of the two-opinions Naming Games 
is applicable to the roles and behaviour of
neutrals in social forums or blogs, and
represent a significant departure from the linguistic roots of the
original NG.
\end{abstract}

\maketitle

\section{Introduction}

The Naming Games have been widely studied as a class of effective
models for information sharing and opinion spread in social
networks. Naming Games were first introduced in linguistics to model
the convergence of social names or tags to a small or even single
label \cite{Baronchelli06}, \cite{Baronchelli08}. In the two
opinions subclass (where the opinions or signals are denoted A and
B) of the NG, the original model treats the neutrals as agents who
have equal probability of signalling A or B. Thus, a speaker -
listener pair chosen from the neutrals can either exit as a pair of
purely A or B opinions agents. The neutral speaker has half
probability of signalling A which is assumed to be received by the
neutral listener with full fidelity who then, along with the
speaker, becomes a purely A opinion agent. The same holds if the
speaker signals the opinion B in which case, both speaker and
listener switch from neutrals to purely B.

One of the previous point of departure from this original NG
\cite{Baronchelli12} is the listener-only version (LO-NG) analyzed
in \cite{Xie11,ZhangLim} where the asymmetry between speaker and
listener is changed significantly to empower the speaker in the
sense that only the listener may switch to a different opinion type
in a single time step. This was done at first to simplify the
mathematical analysis of the NG but we emphasize another important
aspect of the LO-NG, namely, that it more closely resemble social
opinion dynamics and information sharing and less the original
linguistic applications of the NG. Arguably, the agent that speaks
is the more pro-active of the pair and seeks to convince the other
of his opinions.

One aim in this paper is to analyze the nuances of the root concept
of \emph{neutral} in sociology settings where the above speaker -
listener protocol have a range of asymmetry. We focus on three
related interpretations of this concept, namely, (a)
\emph{undecided} as in binary elections or referendums to choose
between two candidates A, B, (b) \emph{middle-roaders} as in blogs
or social forums on a single issue with FOR and AGAINST as the
extreme opinions and possibly several intermediate opinions in
between them, such as in the issue of gun control, and (c)
\emph{erratic or flippant} as in the listener only (LO-NG) versions
of the two-opinions NG where the neutrals or AB agents speak the
either opinion A or B with fixed probabilities.

Clearly, for category (a) the undecided neutrals (AB) will be more
reluctant to speak than the decided agents (A or B), and this could
be formalized by neutrals never chosen as speakers, only as
listeners in the associated game. Otherwise, the interactions are
the same as in the LO-NG.

In category (b) the middle-roaders in social forums say, are neither
undecided nor flippant, but rather, are very likely to be staunch
believers in their middle opinion or opinions (AB). This neutral
behaviour can be formalized by what we call the M models, where the
agents are divided into two disjoint classes, namely, (1) the
extremers who hold the extreme opinions (A or B), convert other
extremers to the opposite extreme opinion, and are converted to
middle-roaders only by (single or multiple) exposure to the Devil's
Advocacy of middle-roaders given next, and (2) the middle-roader
neutrals who do not interact with other neutrals, play the Devil's
Advocate when speaking to extremers ( eg, speak B to an A extremer
or vice-versa), and when listening to an A(resp. B) extremer (once
or multiple times) converts to the corresponding extreme opinion.

To put (b) into a wider context, we add for completeness another
category (c) of middle-roader neutral behaviour who not only play
the Devil's Advocate when speaking to extremers, but when listening
once or more times to an A (resp. B) extremer converts to the other
extreme opinion B (resp. A).

The original LO-NG fits into this classification in the category (d)
of neutrals (AB) who are flippant (randomly speak A or B) when
speaking to both other neutrals and extremers, and when listening,
converts to the extremer type corresponding to the signal received.
In this same category, the extremers consistently speak their
associated opinion, and when listening, convert to flippant neutrals
upon receiving the opposite extreme opinion.

Finally, again for the sake of being more complete, we
describe the category (e) which differs from the flippant neutral (LO-NG) 
only in the perverse conversion of a neutral (AB) to the
opposite B (resp. A) extremer when receiving signal A (resp. B),
irrespective of the speaker.

There are more categories and the complete list can be tabulated
according to different versions of the two-body three urns models
\cite{pickering}. It seems that only the case
(b) of middle-roaders can be solved exactly in the sense of
diagonalizing the Markov transition matrix in closed form by using
the generating functions method pioneered by Kac in his solution of
the Ehrenfest Urn Problem \cite{ehrenfest}. Moreover, as shown below it is
equivalent to the 3-Voter model.

In the rest of the paper, we will solve the \emph{M-models} exactly
to compute for large network1s, expected times to consensus. The
results obtained here, without using the Diffusion Approximation or
Fokker-Planck Master equation, but rather from direct probabilistic
calculations at the level of random walks, agree with those in
\cite{Santorini12}. Moreover, our calculations below can be extended
for all values of $M$ including the smallest case $M=3$ which is of
special interest here because of its equivalence to the scenario (b)
of middle-roaders neutrals in social dialog and blogs.

The $M$ models will be solved exactly to prove the result that
starting from a population of $N/3$ agents in any one of the $M$
microstates including the $M-2$ neutral ones, the expected times to
consensus, $\tau = O(N^2)$ is much longer than the
$O(NlogN)$ expected consensus times for the two-opinions LO-NG model
in scenario (c) for erratic neutrals \cite{Xie11}, \cite{ZhangLim}.

\section{Random Walk Model}
We assume that the model is imposed on a complete graph. That is, every node is connected to every other node in the network. Let $n_A(m)$, $n_B(m)$, and $n_C(m)$ be the total number of nodes with opinions $A$, $B$, and $C$ respectively at discrete time step $m$. Since the total number of nodes in the graph, $N$, is constant, we can eliminate $n_C$ by taking $n_C(m)=N-n_A(m)-n_B(m)$. So, the probability distribution of macrostates $(n_A(m)=i, n_B(m)=j)$ at time step $m+1$ can be expressed in terms of information given at time $m$ as
\begin{equation}
\begin{split}
a_{ij}^{(m+1)}=&p^{A}_{i-1j}a_{i-1j}^{(m)}+p^{A}_{i+1j}a_{i+1j}^{(m)}+p^{B}_{ij-1}a_{ij-1}^{(m)}+p^{B}_{ij+1}a_{ij+1}^{(m)}\\
&+p^{A-B}_{i-1j+1}a_{i-1j+1}^{(m)}+p^{A-B}_{i+1j-1}a_{i+1j-1}^{(m)}+r_{ij}a_{ij}^{(m)}.\label{1propagator}
\end{split}
\end{equation}
where
\begin{equation}
\begin{tabular}{ll}
$p^{A}_{ij}=q^{A}_{ij}=\frac{i(N-i-j)}{N(N-1)}$ & $p^{B}_{ij}=q^{B}_{ij}=\frac{j(N-i-j)}{N(N-1)}$\\
$p^{A-B}_{ij}=q^{A-B}_{ij}=\frac{ij}{N(N-1)}$ & $r_{ij}=1-2p^{A}_{ij}-2p^{A-B}_{ij}$.
\end{tabular}
\end{equation}
Each pair $(i,j)$ in equation \eqref{1propagator} corresponds to a row of the Markov transition matrix, $\mathbf{T}$. The $m$ step propagator, $a_{ij}^{(m)}$, can be found by multiplying the initial distribution, $a_{ij}^{(0)}$, in vector form by the matrix $\mathbf{T}^m$. We will find the $m$ step propagator analytically by solving the spectral problem for the matrix $\mathbf{T}$.

\section{The Spectral Problem}
We extend the procedure found in ref. \cite{pickering} for the 3 opinion Voter model on the complete graph. For eigenvalue $\lambda$ and corresponding eigenvector $\mathbf{v}$ with components $c_{ij}$, the spectral problem can be stated as
\begin{equation}
\begin{split}
(\lambda-1) c_{ij}=&p^{A}_{i-1j}c_{i-1j}+p^{A}_{i+1j}c_{i+1j}+p^{B}_{ij-1}c_{ij-1}+p^{B}_{ij+1}c_{ij+1}+p^{A-B}_{i-1j+1}c_{i-1j+1}\\
&+p^{A-B}_{i+1j-1}c_{i+1j-1}-(2p_{ij}^A+2p_{ij}^B+2p_{ij}^{A-B})c_{ij}.\label{spectral}
\end{split}
\end{equation}
To solve this, we introduce a generating function $G(x,y,z)$ defined by
\begin{equation}
G(x,y,z)=\sum_i\sum_jc_{ij}x^iy^jz^{N-i-j}.\label{genfun}
\end{equation}
The double sum is over all pairs $(i,j)\in\mathbb{Z}^2$, however we force the restriction that $c_{ij}=0$ when $i<0$, $j<0$ or $i+j>N$. These restrictions are equivalent to the requirement that there cannot be any negative powers in $G$. Using these, we express the spectral problem in equation \eqref{spectral} as a PDE in generating function form:
\begin{equation}
(x-z)^2G_{xz}+(y-z)^2G_{yz}+(x-y)^2G_{xy}=N(N-1)(\lambda-1)G.
\end{equation}
To solve this, make the change of variables $u=x-y$, $v=y-z$, and $H(u,v,z)=G(x,y,z)$. Since this is a linear change of variables, we expect $H$ to have the same form as $G$, except with coefficients $b_{ij}$. The PDE for the spectral problem expressed in the new variables becomes
\begin{equation}
(u^2+2uv)H_{uz}-2uvH_{uv}+v^2H_{vz}-v^2H_{vv}-u^2H_{uu}=N(N-1)(\lambda-1)H.
\end{equation}
This difference equation for the coefficients of $H$ can be written explicitly as
\begin{equation}
b_{ij}=\frac{(N-i-j+1)[(i-1)b_{i-1j}+(j-1)b_{ij-1}]}{N(N-1)(\lambda-1)+(i+j)(i+j-1)}\label{explicitform}.
\end{equation}
To find the eigenvalues and eigenvectors, we only consider the cases in which equation \eqref{explicitform} becomes singular for some $i=i'$, $j=j'$. This is possible only when the denominator in equation \eqref{explicitform} is zero at this point. This shows that for $i',j'=0...N,\; i'+j'\leq N$, the set of eigenvalues is
\begin{equation}
\lambda_{i'j'}=1-\frac{(i'+j')(i'+j'-1)}{N(N-1)}\label{eigenvalues}.
\end{equation}
If we take $\lambda=\lambda_{i'j'}$, then $b_{i'j'}$ can take any value. This is a reflection of the fact that any multiple of an eigenvector remains an eigenvector with the same eigenvalue. Without loss in generality, take $b_{i'j'}=1$. Now, for $i>i'$, and $j>j'$, equation \eqref{explicitform} will never be singular and non-trivial. This will determine all of the values of $b_{ij}$ for any $\lambda_{i'j'}$.

Now we find $c_{ij}$ in terms of $b_{ij}$. We do this by expressing $H(u,v,z)$ in the original $(x,y,z)$ coordinates to obtain
\begin{equation}
c_{ij}=\sum_{k=i}^{i+j}\sum_{r=i+j-k}^{N-k}b_{kr}{k\choose i}{r\choose {i+j-k}}(-1)^{r-j}\label{eigenvectors}.
\end{equation}
Thus we have found the solution to all eigenvalues and eigenvectors explicitly. Note that even though many eigenvalues are repeated, we find independent expressions for $b_{ij}$ for each repetition. This will give independent expressions for $c_{ij}$ for each independent $b_{ij}$ that had been computed above. This suggests that the matrix is diagonalizable, which allows us to express any distribution in the eigenbasis of $\mathbf{T}$.

\section{Applications of the Spectral Solution}
The solution to the spectral problem can be used to find many useful quantities related to the 3-Voter model. This is because we now know the $m$ step propagator exactly. We can express the probability of each macrostate as:

\begin{equation}
a_{ij}^{(m)}=\sum_{i'j'} d_{i'j'}\lambda_{i'j'}^m[\mathbf{v_{i'j'}}]_{ij}\label{mpropagator}
\end{equation}
Here, $[\mathbf{v_{i'j'}}]_{ij}$ is the $i,j$ component of the eigenvector corresponding to the eigenvalue $\lambda_{i'j'}$. The coefficients $d_{i'j'}$ represent the initial macrostate probability distribution, $a_{ij}^{(0)}$, expressed in the eigenbasis. That is, we gather the eigenvectors into a matrix $\mathbf{V}$ and solve $\mathbf{Vd}=\mathbf{a}^{(0)}$, where the components of $\mathbf{d}$ and $\mathbf{a}^{(0)}$ are $d_{i'j'}$ and $a_{ij}^{(0)}$ respectively. Since $\mathbf{T}$ is diagonalizable, $\mathbf{V}$ is invertible. This form is very convenient in the applications that follow.

\subsection{Moments of Consensus Time}
With the solution to the spectral problem known, we can find all moments of the consensus time exactly. The consensus time is the number of steps required until every node in the network has the same opinion state. Once the model has reached consensus, the system will never leave this state. In addition to the exact solution for the moments, we will also find an estimate of this quantity that is independent of the initial distribution that demonstrates the asymptotic dependence on $N$ and the moment, $p$.

Let $q_m$ be the probability that the system reaches consensus at time $m$. The $p^{th}$ moment of consensus time, $\tau$, as a function of the initial distribution can expressed as
\begin{equation}
E[\tau^p|\mathbf{d}=\mathbf{V^{-1}a}^{(0)}]=\sum_{m=0}^\infty q_m m^p.\label{consensusmoments}
\end{equation}
We find the explicit form for $q_m$ to be
\begin{multline}
q_m=\frac{1}{N}[a_{1,0}^{(m-1)}+a_{0,1}^{(m-1)}+a_{N-1,0}^{(m-1)}+a_{N-1,1}^{(m-1)}+a_{0,N-1}^{(m-1)}+a_{1,N-1}^{(m-1)}].\label{qm}
\end{multline}
Apply equation \eqref{mpropagator} to express this as

\begin{multline}
q_m=\frac{1}{N}\mathop{\sum_{i'=0}^N\sum_{j'=0}^{N-i'}}_{i'+j'>1} d_{i'j'}\lambda_{i'j'}^{m-1}\{[\mathbf{v_{i'j'}}]_{1,0}+[\mathbf{v_{i'j'}}]_{0,1}\\
+[\mathbf{v_{i'j'}}]_{N-1,0}+[\mathbf{v_{i'j'}}]_{N-1,1}+[\mathbf{v_{i'j'}}]_{0,N-1}+[\mathbf{v_{i'j'}}]_{1,N-1}\}.
\end{multline}
The sum is taken so that $i'+j'>1$ since the macrostate probabilities in equation \eqref{qm} are independent of the other eigenvectors. That is, $d_{i'j'}=0$ for $i'+j'\leq0$. This is because $\lambda_{i'j'}=1$ in these cases, which corresponds to the consensus points. To simplify notation, we define $s_{i'j'}$ to be
\begin{multline}
s_{i'j'}=d_{i'j'}\{[\mathbf{v_{i'j'}}]_{1,0}+[\mathbf{v_{i'j'}}]_{0,1}+[\mathbf{v_{i'j'}}]_{N-1,0}\\
+[\mathbf{v_{i'j'}}]_{N-1,1}+[\mathbf{v_{i'j'}}]_{0,N-1}+[\mathbf{v_{i'j'}}]_{1,N-1}\}\label{sij}.
\end{multline}
Therefore, we express the moments of consensus time as
\begin{align}
E[\tau^p|\mathbf{d}=\mathbf{V^{-1}a}^{(0)}]\sim \frac{1}{N}\mathop{\sum_{i'=0}^N\sum_{j'=0}^{N-i'}}_{i'+j'>1} s_{i'j'}\frac{p!}{(1-\lambda_{i'j'})^{p+1}}.
\end{align}
This form for the moments of consensus time is a function of the initial distribution through the variable $s_{i'j'}$ defined in equation \eqref{sij}. The moments of consensus time usually vary by an $O(1)$ multiple when the initial distribution changes. Only when the system is initially near consensus will this multiple vary with $N$. If we assume this is not the case, we now seek to find the asymptotic dependence of the moments of consensus time on $N$ and the moment $p$.

Let $S^{(m)}$ be the probability that the system is not in consensus at time $m$. This is the sum over all probabilities $a_{ij}^{(m)}$ that do not correspond to consensus states. These probabilities are independent of the eigenvectors with eigenvalue $\lambda=1$ since consensus is a frozen state. Thus, $S^{(m)}=O(\lambda_2^m)$, where $\lambda_2$ is the second largest eigenvalue. Therefore, the probability of entering consensus at time $m$ is given by $q_m=S^{(m-1)}-S^{(m)}=O\left(\lambda_2^m\frac{1-\lambda_2}{\lambda_2})\right)$. Now, we substitute this into equation \eqref{consensusmoments} to obtain

\begin{equation}
E[\tau^p]=O(p!2^{-p}N^{2p})
\end{equation}

\begin{figure}[h!]
\begin{center}
\includegraphics[scale=.33]{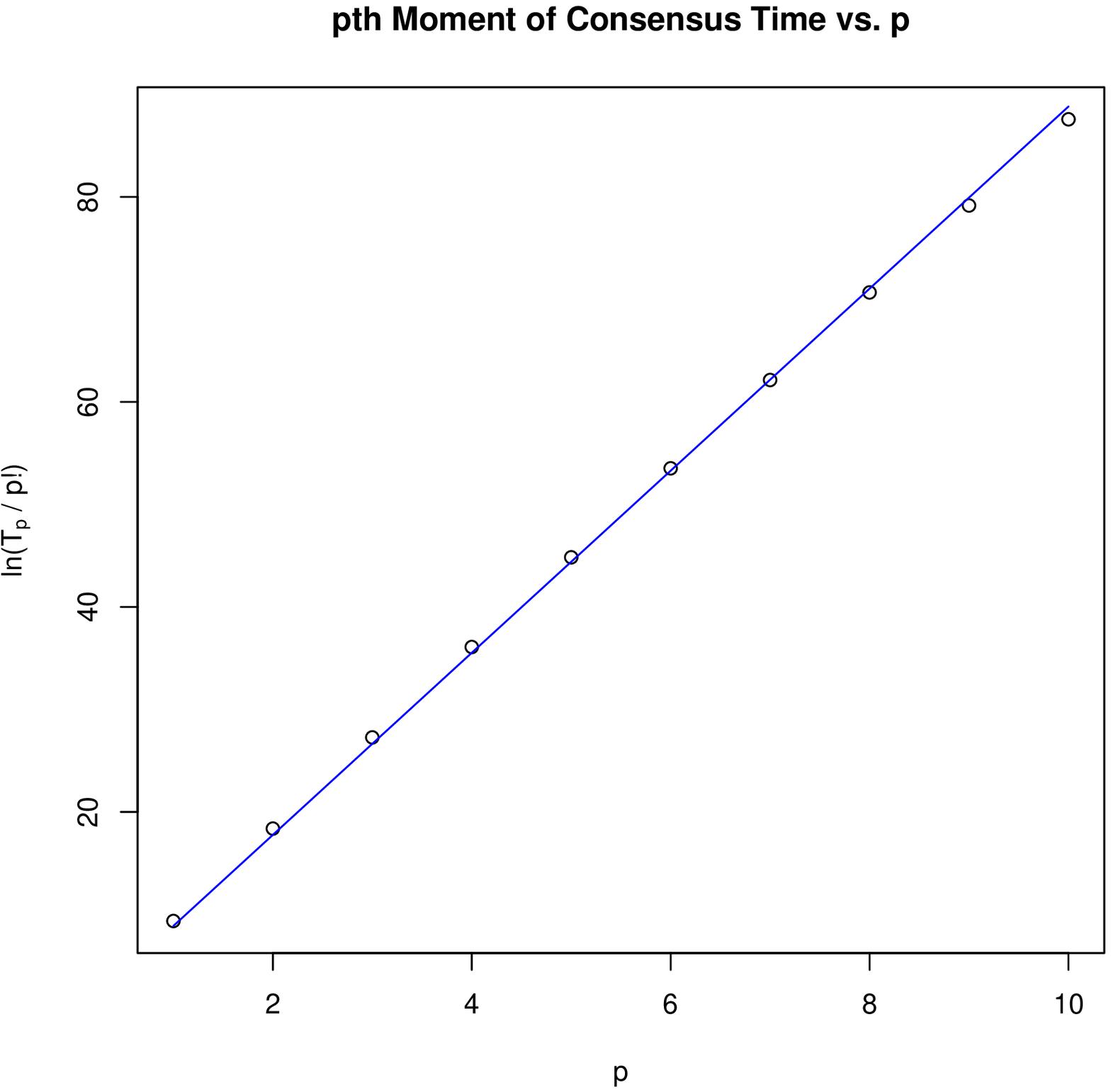}
\includegraphics[scale=.33]{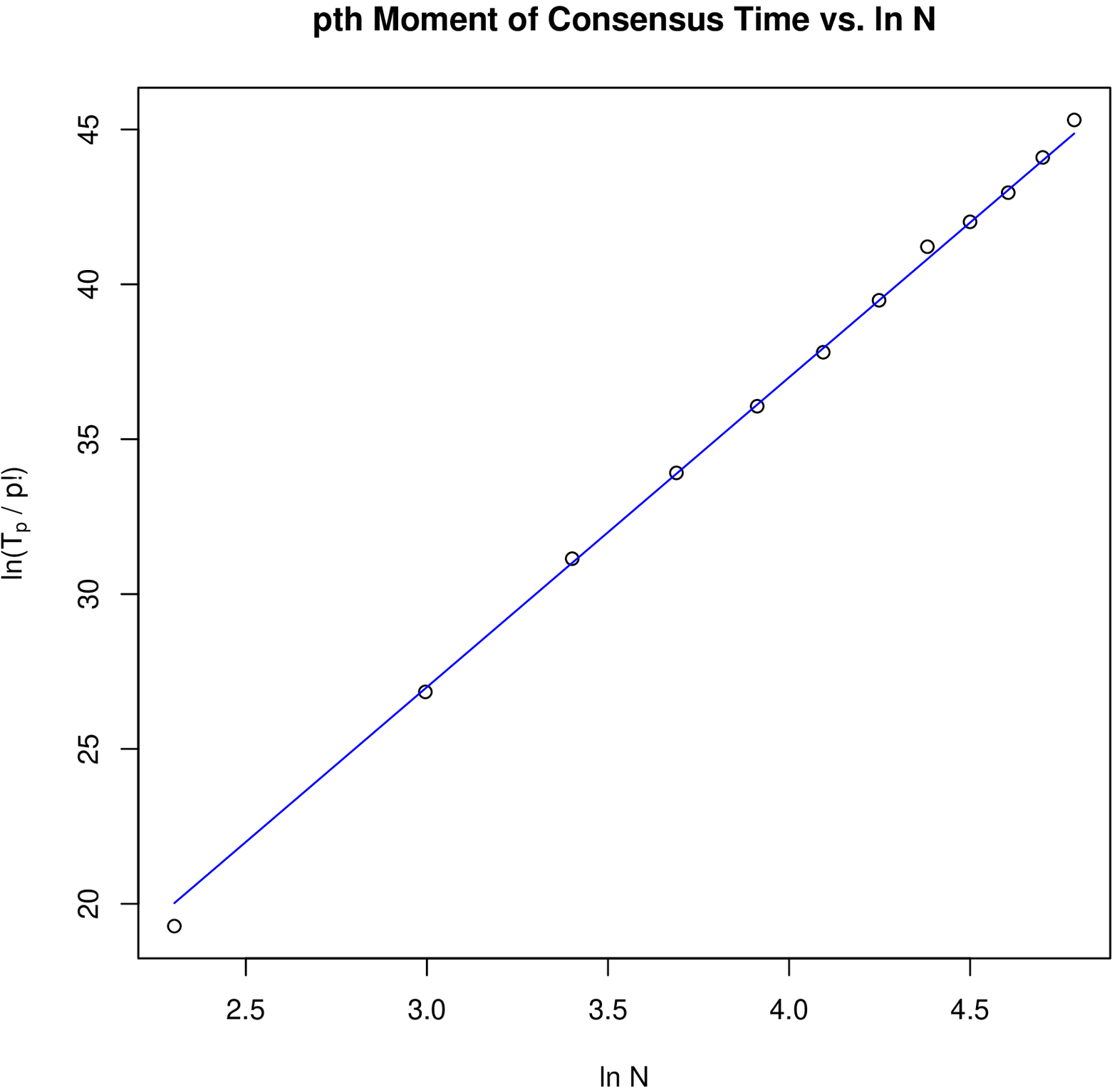}.
\end{center}
\caption{Monte carlo simulation results for the dependence on $p$ and $N$. In the top figure, we fix $N=120$ and take $p=1...10$. A line with slope $2\ln N-\ln2$ corresponding to the theoretical prediction is also plotted. In the bottom figure, we fix $p=5$ and let $N=10,20,...,120$. The line is the best fit of the data that has slope $10$, which is the theoretical result we expect. These results show good correspondence between simulated and theoretical results.}\label{consensusfig}
\end{figure}

This estimate is uniform over all initial distributions and considered a function of $p$ and $N$ only. This result shows that the expected time to consensus is $O(N^2)$. We reinforce this result with two Monte Carlo simulations. In the first simulation, we fix $N=120$ and run the 3 opinion Voter model over 3000 runs. Let $T_p$ be the average of $\tau^p$ over all runs, which is the estimate of the $pth$ moment of consensus time. We predict that there is a linear relationship between $\ln(T_p/p!)$ and $p$, and that the slope of this line is about $2\ln N-\ln2$. In the second simulation, we fix $p=5$ and let $N$ vary, we expect to find a linear relationship between $\ln (T_p/p!)$ and $\ln N$ with slope $2p=10$. Figure \ref{consensusfig} shows the comparison between the simulated and predicted values.

\subsection{Local Times}
We can also apply the solution of the spectral problem to find all local times of the 3-Voter model. We define local times to be the expected time spent at each macrostate $(n_A=i, n_B=j)$ prior to consensus. We only consider macrostates that do not correspond to consensus states. Local times are stronger quantities than consensus times, because the sum of all local times will be equivalent to the expected time to consensus.

Let $M_{ij}(m)$ be the number of times steps spent at macrostate $(n_A=i, n_B=j)$ by time $m$. This is a monotonically increasing random walk, where
\begin{equation}
M_{ij}(m+1)=M_{ij}(m)+\Delta M_{ij}(m).
\end{equation}
\begin{figure}[h!]
\begin{center}
\includegraphics[scale=.4]{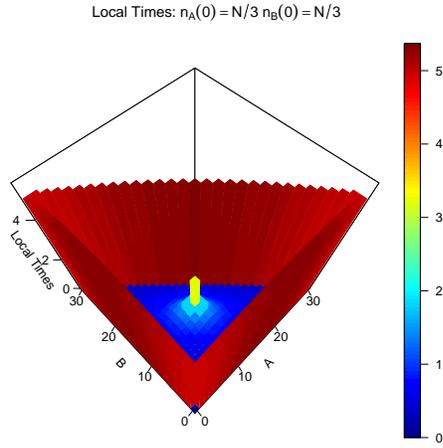}
\end{center}
\caption{Local times of the 3-Voter model in the case when $n_A(0)=N/3$, $n_B(0)=N/3$ and $N=30$. The model spends most of the time at the boundary, in which the model is reduced to the 2-Voter model.}\label{localtime}
\end{figure}
At time $m$, the randomness of the walk is exhibited in $\Delta M_{ij}(m)$, which can take values in $\{0,1\}$. If the model is in macrostate $(n_A=i, n_B=j)$ at time $m+1$, then $\Delta M_{ij}(m)=1$. This occurs with probability $a_{ij}^{(m+1)}$. So, we have that $E[\Delta M_{ij}(m)]=a_{ij}^{(m+1)}$. With this, we can find the local time for macrostate $(n_A=i, n_B=j)$:
\begin{align}
E[M_{ij}(\infty)-M_{ij}(0)]=\sum_{m=0}^\infty a_{ij}^{(m)}
\end{align}
Utilize equation \eqref{mpropagator} to obtain the exact expression for local times:
\begin{align}
E[M_{ij}(\infty)]=N(N-1)\mathop{\sum_{i'=0}^N\sum_{j'=0}^{N-i'}}_{i'+j'>1}\frac{d_{i'j'}[\mathbf{v_{i'j'}}]_{ij}}{(i'+j')(i'+j'-1)}.
\end{align}
We write the sum such that $i'+j'>1$ because we do not consider consensus states when computing local times. The macrostates we consider are independent of the eigenvalue $\lambda=1$, which correspond to $i'+j'=0$ or $i'+j'=1$. Figure \ref{localtime} shows the local time for the 3-Voter model when $n_A(0)=N/3$, $n_B(0)=N/3$ and $N=30$.

\section{Conclusion}
We have found concrete mathematical correlates for three different
interpretations of neutral in sociology and politics. All three
mathematical models belong to the generalized Listener Only NG and
can therefore be systematically analyzed for key statistical
quantities such as expected times to multi-consensus as in this
paper.

In general, the deterministic drift in the coarse-grained or random
walk models corresponding to each of these three families of NG-like
models plays a key role. With increasing slow-time drift towards
points of consensus, the expected times to consensus are decreased
for most initial states. Meanwhile, such an increased drift against
the committed minority consensus (near the consensus point of the
position without committed fraction) generally predicts a relatively
larger tipping fraction of committed minority agents, as a direct
mathematical consequence of the saddle-node bifurcation on the
slow-time manifold in phase-space.

Specifically, we have given exact solutions for expected times to
multi-consensus $\tau$ in the \emph{M-models} for all values of
$M>2$ - $\tau\approx O(N^2)$. Compared to the corresponding times
for the original LO-NG on two opinions in scenario (c) which is
$O(N\ln N)$, the strong neutrals (middle-roaders) have, as predicted
in an early section, increased the expected times to multi-consensus
for large networks. For a complete list of cases of the 2-Voter models 
solvable by this generating function
approach, we refer the reader to a prior paper \cite{pickering}.

Another conclusion that we can draw from the drift-less nature of
the $M$ models is that there are no positive (nonzero) tipping
points for committed minority agents in any opinion type - even a
very small number of committed agents of any type will tilt the $M$
game dynamics towards faster consensus of that opinion. This can be
viewed as a disadvantage of the $M$ models for scenario (b) because
bloggers in a social forum of different neutral positions are
unlikely to be bowled over by extremely small numbers of committed
leader-agents. In future work, we will attempt to fix this defect.

\section*{Acknowledgements}
This work was supported in part by the
Army Research Office Grants No. W911NF-09-1-0254 and
W911NF-12-1-0546. The views and conclusions contained in this
document are those of the authors and should not be interpreted as
representing the official policies, either expressed or implied, of
the Army Research Office or the U.S. Government.

\section*{References}
\bibliography{3voterb}

\end{document}